\documentclass[]{rcdj}

\begin{document}

\setcounter{page}{101}
\journal{REGULAR AND CHAOTIC DYNAMICS, V.\,9, \No2, 2004}
\runningtitle{ABSOLUTE AND RELATIVE CHOREOGRAPHIES}
\title{ABSOLUTE AND RELATIVE CHOREOGRAPHIES IN THE PROBLEM OF
POINT VORTICES MOVING ON A PLANE}
\runningauthor{A.\,V.\,BORISOV, I.\,S.\,MAMAEV, A.\,A.\,KILIN}
\authors{A.\,V.\,BORISOV, I.\,S.\,MAMAEV, A.\,A.\,KILIN}
{Institute of Computer Science\\
1, Universitetskaya str.\\
426034, Izhevsk, Russia\\
E-mail: borisov@rcd.ru\\
E-mail: mamaev@rcd.ru\\
E-mail: aka@rcd.ru}

\abstract{We obtained new periodic solutions in the problems of three and four point
vortices moving on a plane. In the case of three vortices, the system is
reduced to a Hamiltonian system with one degree of freedom, and it is
integrable. In the case of four vortices, the order is reduced to two
degrees of freedom, and the system is not integrable. We present
relative and absolute choreographies of three and four  vortices of
the same intensity which are periodic motions of vortices in some
rotating and fixed frame of reference, where all the vortices move along
the same closed curve. Similar choreographies have been recently obtained
by C.\,Moore, A.\,Chenciner, and C.\,Simo for the~$n$-body problem in
celestial mechanics~\cite{ChencMontg,ChencGervMontgSim,Simo}. Nevertheless, the choreographies that appear in
vortex dynamics have a number of distinct features.}
\amsmsc{76B47, 37J35, 70E40}
\doi{10.1070/RD2004v009n02ABEH000269}
\received 05.04.2004.

\maketitle

\paragraph{Equations of motion and first integrals.}
Let us review briefly basic forms of equations and first integrals in the
dynamics of point vortices on a plane (for a detailed discussion,
see~\mbox{\cite{bmk-1,bmk-5,bmk-2}}, where, in addition, hydrodynamical
assumptions required for the validity of these equations are specified).

\paragraph{}
For $n$ point vortices with Cartesian coordinates~$(x_i,y_i)$ and
intensities~$\Gamma_i$, the equations of motion can be written in
Hamiltonian form,
\begin{equation}
\label{bmk-eq-1} \Gamma_i \dot x_i=\pt{\Cal H}{y_i},\quad \Gamma_i \dot
y_i=-\pt{\Cal H}{x_i},\quad 1\le i\le n,
\end{equation}
where the Hamiltonian  is
\begin{equation}
\label{bmk-eq-2}
{\Cal H}=-\frac{1}{4\pi}\sum_{i<j}^n{}\Gamma_i\Gamma_j\ln|\bs r_i-\bs
r_j|^2,\quad\bs r_i=(x_i,y_i).
\end{equation}
Here, the Poisson bracket is:
\begin{equation}
\label{bmk-eq-5}
\{f,g\}=\sum_{i=1}^N\frac{1}{\Gamma_i}\Bigl(\pt{f}{x_i}\pt{g}{y_i}-\pt{f}{y_i}\pt{g}{x_i}\Bigr).
\end{equation}
Equations \eqref{bmk-eq-1} are invariant under the action of the Eucledian group
$E(2)$, so (in addition to the Hamiltonian) they also have three first
integrals
\begin{equation}
\label{bmk-eq-3}
Q=\sum_{i=1}^n\Gamma_i x_i,\quad P=\sum_{i=1}^n \Gamma_i y_i,\quad
I=\sum_{i=1}^n \Gamma_i(x_i^2+y_i^2),
\end{equation}
which, however, are not involutive,
\begin{equation}
\label{bmk-eq-4}
\{Q,P\}=\sum_{i=1}^N\Gamma_i,\quad \{P,I\}=-2Q,\quad \{Q,I\}=2P.
\end{equation}

Using these three integrals, it is possible to construct two involutive
integrals, $Q^2+P^2$ and~$I$. Hence, according to the general
theory~\cite{bmk-6, MarsdWeinst}, we can reduce the number of degrees of
freedom by two. Thus, in the particular case of three vortices, the system
can be reduced to one degree of freedom and becomes integrable (Gr\"obli,
Kirchhoff, Poincar\'{e})~\cite{bmk-1,bmk-5,bmk-2,Synge}, while the problem of
four vortices is reduced to a system with two degrees of freedom.
Generally, the last problem is not integrable \cite{bmk-7}.

Effective reduction in the system of four vortices with intensities of
the same sign was done by K.\,M.\,Khanin in~\cite{bmk-4}. In that work two pairs of vortices were considered and  for each
pair action-angle variables were selected.  The
system of four vortices was then obtained as a perturbation of the two two-vortex systems.  He
proves (using the methods of KAM-theory) the existence of quasiperiodic
solutions. As a small parameter, he takes a value inverse to the distance
between two pairs of vortices.

Reduction by one degree of freedom using the translational invariants~$P$
and~$Q$ was done by Lim in~\cite{bmk-8}. He introduced barycentric Jacobi
coordinates (centered, in this case, at the center of vorticity), which
have well-known analogs in the classical~$n$-body problem in celestial
mechanics~\cite{bmk-3}. Note that even this (partial) reduction made it
possible to apply some methods of KAM-theory to the problem of point
vortices' motion~\cite{bmk-8,Lim-2}.

When $n$ vortices on a plane have equal intensities, one of the more
formal methods of order reduction, which was described in~\cite{bmk-9}
(see also~\cite{bmk-10,bmk-1}), proved to be most suitable. This method is
based on using mutual variables representation of the equations of motion. As mutual variables, the
squares of distances between the pairs of vortices and oriented areas of
triangles were taken:
\begin{equation}
\label{D01}
M_{ij}=(x_i-x_j)^2+(y_i-y_j)^2,\quad\Delta_{ijk}=(\bs r_j-\bs
r_k)\wedge(\bs r_k-\bs r_i).
\end{equation}
Mutual commutation of such variables (which are introduced by E.\,Laura)
leads to some Lie algebra. Here, the reduction procedure (more exactly,
the last canonical step of it) is equivalent to a purely algebraic problem
of introduction of symplectic coordinates on the orbits of corresponding
Lie algebras.
(For identical vortices the method of reduction based on the Fourier transformatio was introduced also
in~\cite{ArefPomp}.)

\paragraph{Reduction for three and four vortices of equal
intensity.} Without losing in generality, we put~$\Gamma_i=\Gamma_j=1$,
$P=Q=0$, then the moment integral~$I$ \eqref{bmk-eq-3} can be written as
\begin{equation}
\label{D1}
I=\frac1n\sum\limits_{i<j}^nM_{ij},
\end{equation}
where $n$ is the number of vortices. Using the complex representation of
the vortices' coordinates, $z_k=x_k+iy_k$, we obtain:
\begin{equation}
\label{D2}
z_k=\frac1n\sum\limits_{j=k}^n\sqrt{M_{kj}}e^{i\theta_{kj}},
\end{equation}
where $\theta_{kj}$ is the angle between the vector from the~$j$-th vortex
to the~$k$-th vortex and the positive direction of $Ox$-axis.

The following propositions, defining the dynamics of a reduced system of
vortices, can be obtained by direct computation:

\begin{pro}\label{pro1}
For three vortices of equal intensity, evolution of the mutual distances~\eqref{D01}
$($assuming~$I=\const)$ is described by a Hamiltonian system with one
degree of freedom. In canonical variables~$(g,G)$, the system looks as
\begin{equation}
\label{D3}
\dot g=\frac{\partial{\Cal H}}{\partial G},\quad
\dot G=-\frac{\partial{\Cal H}}{\partial g},\quad
{\Cal H}=-\frac1{4\pi}\ln M_{12}M_{13}M_{23},
\end{equation}
where $M_{12}=4\left(\frac I2-G\right)$, $M_{13}=8G-I+2\sqrt{12}
\sqrt{\left(\frac I2-G\right)G}\cos g$, $M_{23}=4\left(\frac
I2-G\right)-2\sqrt{12} \sqrt{\left(\frac I2-G\right)G}\cos g$.
\end{pro}

\begin{pro}\label{pro2}
For four vortices of equal intensity, evolution of the mutual distances is
described by a Hamiltonian system with two degrees of freedom. In
canonical variables $(g,G,h,H)$, the system looks as
\begin{gather}
\label{D4}
\dot g=\frac{\partial{\Cal H}}{\partial G},\quad
\dot G=-\frac{\partial{\Cal H}}{\partial g},\quad
\dot h=\frac{\partial{\Cal H}}{\partial H},\quad
\dot H=-\frac{\partial{\Cal H}}{\partial h},\\
{\Cal H}=-\frac1{4\pi}\ln M_{12}M_{13}M_{14}M_{23}M_{24}M_{34},
\notag
\end{gather}
where
$$
\begin{aligned}
M_{12}&=I-G+2\sqrt{(I-H)(I-G)}\cos h,&\quad
M_{34}&=I-G-2\sqrt{(I-H)(I-G)}\cos h,\\
M_{13}&=I+G+2\sqrt{(I-H)G}\cos(h+g),&\quad
M_{24}&=I+G-H-2\sqrt{(I-H)G}\cos(h+g),\\
M_{14}&=H+2\sqrt{(H-G)G}\cos g,&\quad
M_{23}&=H-2\sqrt{(I-G)G}\cos g.
\end{aligned}
$$
\end{pro}

\begin{rem*}
These canonical variables are of natural geometrical origin related to
representation of the equations of motion on a Lie algebra
\cite{bmk-9,bmk-10}.
\end{rem*}

\paragraph{Absolute motion: quadratures and geometric interpretation.}
According to \eqref{D2}, when~$M_{ij}(t)$ are known, one needs to know the
angles~$\theta_{ij}(t)$ to determine the vortices' coordinates. It is
obvious that only one of the angles is independent (in this case, we
take~$\theta_{12}$ to be independent), the remaining angles are computed
with the cosine theorem:
\begin{equation}
\label{star1}
\theta_{ij}+\theta_{ik}=\arccos\left(\frac{M_{jk}-M_{ij}-M_{ik}}{2\sqrt{2M_{ij}M_{ik}}}\right),
\quad i\ne j,\;k\ne i.
\end{equation}
Evolution of $\theta_{12}$ is obtained by quadrature~\cite{bmk-5}:
\begin{equation}
\label{star}
\begin{aligned}
4\pi\dot\theta_{12}&=M_{13}^{-1}+M_{23}^{-1}+M_{12}^{-1}(6-M_{23}^{-1}M_{13}-M_{13}M_{23}^{-1})\quad(\text{for
three vortices}),\\
4\pi\dot\theta_{12}&=M_{13}^{-1}+M_{14}^{-1}+M_{23}^{-1}+M_{24}^{-1}+\\
{}&\quad+M_{12}^{-1}(8-M_{13}^{-1}M_{23}-M_{14}^{-1}M_{24}-M_{23}^{-1}M_{13}-M_{24}^{-1}M_{14})
\quad(\text{for four vortices}).
\end{aligned}
\end{equation}

In the case of periodic solutions of the reduced system \eqref{D3},
\eqref{D4}, there is an interesting geometric interpretation of the absolute
motion.

\fig<bb=0 0 131.2mm 83.8mm>{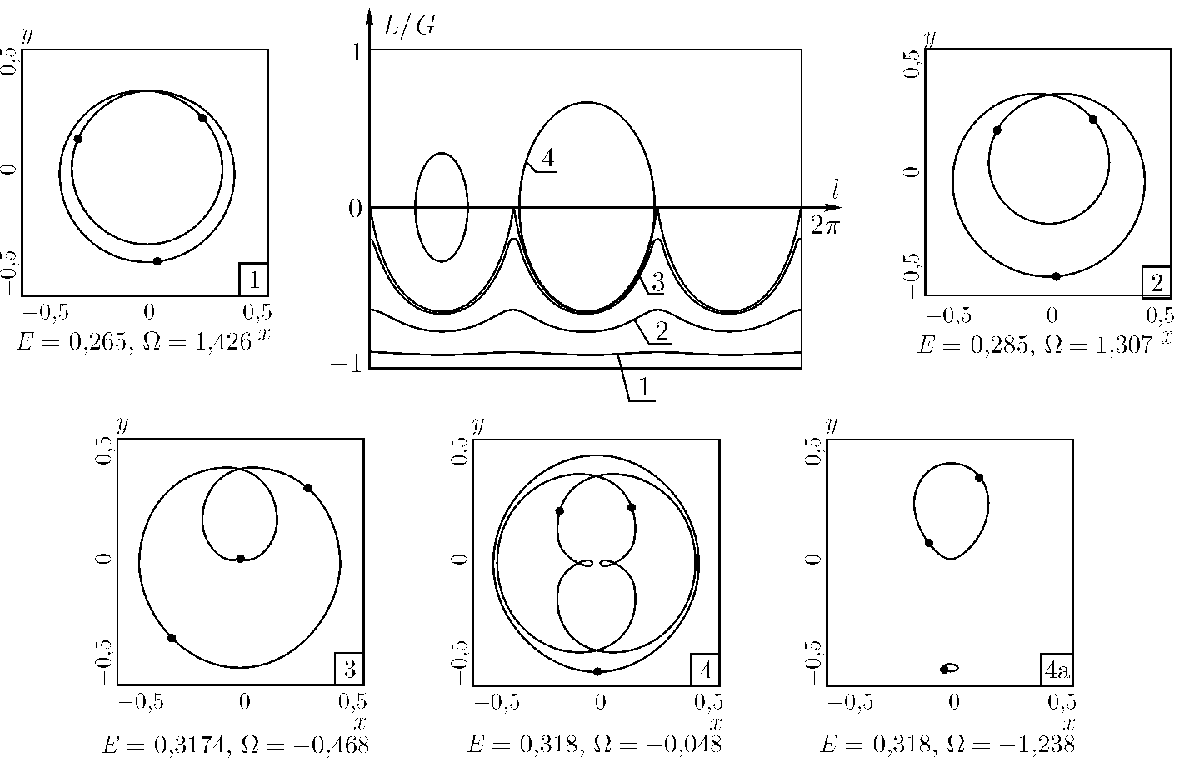}[\label{plane.eps} The phase
portrait of the reduced system in the case of three vortices. Relative
choreographies corresponding to different phase trajectories on the
portrait\vspace{-2mm}]

\begin{pro}\label{pro3}
Let $\gamma(t)$ be a periodic solution $($of period $T)$ of the reduced
system, then\/{\rm

$1^\circ$} there exists a frame of reference, uniformly rotating with some
angular velocity~$\Omega_a$ about the center of vorticity in which each
vortex moves along some closed curve $\xi_i(t);$

{\rm$2^\circ$} the rotational velocity~$\Omega_a$ is given by {\rm(}to
within~$\frac{2\pi}T\frac pq$, $p,q\in\mathbb{Z})${\rm:
\begin{equation}
\label{Om1}
\Omega_a=\frac1T\int\limits_0^T\dot\theta_{12}(t)dt.
\end{equation}

$3^\circ$} if the velocities $\Omega_a$ and~$\Omega_o=\frac{2\pi}T$ are
commensurable $($i.\,e. $\frac{\Omega_a}{\Omega_o}=\frac pq$,
$p,q\in\mathbb{Z}),$ then the vortices in the fixed frame also move along
some closed curves\/{\rm;

$4^\circ$} if any of the curves $\xi_i(t)$ can be superimposed by
rotation about the center of vorticity by an angle, commensurable
with~$2\pi$, then there exists a (rotating) frame of reference such that
the corresponding vortices move along the same curve.
\end{pro}

\proof
We expand the periodic function of period~$T$ in the right-hand
side of~\eqref{star} in a convergent Fourier series\vspace{-2mm}
\begin{equation}
\label{wer2}
4\pi\dot\theta_{12}=\sum_{n\in\mathbb{Z}}a^{(n)}e^{i\frac{2\pi n}Tt}.
\end{equation}
Integrating~\eqref{wer2} while bearing~\eqref{star1} in mind, we come to a
conclusion that the angles~$\theta_{ij}$ depend on~$t$ in the following
way:
\begin{equation}
\label{wer3}
\theta_{ij}(t)=\Omega_{ij} t+g_{ij}(t),
\end{equation}
where $\Omega_{ij}=a^{(0)}+\frac{2\pi}{T}\frac{q_{ij}}{p_{ij}}$,
~$q_{ij},\,p_{ij}\in\mathbb{Z}$, while $g_{ij}(t)=g_{ij}(t+T)$ are
$T$-periodic functions of time.

Substituting this into~\eqref{D2}, we see that locations of the
vortices on a plane is given as follows:
\begin{equation}
\label{wer31}
z_k(t)=\frac{Q+iP}{\sum\Gamma_i}+u_k(t)e^{i\Omega t},\q
u_k(t)=u_k(t+T_1)\in\mathbb{C},\q
\Omega=a^{(0)}.
\end{equation}
Hence, {\it in the frame of reference, rotating about the center of
vorticity with angular velocity~$\Omega$, all the vortices move along
closed analytical curves, given by functions~$u_k(t)\in\mathbb{C}$}.

The proof of $2^\circ$, $3^\circ$, $4^\circ$
with~\eqref{wer2}--\eqref{wer31} is evident.\qed

\begin{rem*}
Proposition~\ref{pro3} is generalized without change to the case of
arbitrarily many vortices,~$n$, provided that~$\gamma(t)$ is a periodic
solution of the reduced system with~$2n-2$ degrees of freedom.
 (For a more detailed discussion of reduction, see,
for example,~\cite{bmk-1}.)
\end{rem*}

\paragraph{Analytical choreographies.} Now we show that the four- and
three-vortex problems can have remarkable periodic solutions, when all the
vortices follow each other along the same curve; such solutions are
referred to as simple (or connected) choreographies. To underline the
difference between choreographies in a fixed frame of reference and
choreographies in a rotating one, these choreographies are
called {\it absolute} and {\it relative}~\cite{ChencGervMontgSim}.

\begin{teo}[\cite{bmk-1}]\label{teo1}
If in the problem of three vortices of equal intensity, the constants of the
integrals of motion,~$I$ and~$\Cal H$, satisfy the inequality
\begin{equation}
\label{D5}
-\ln 3<\frac{4\pi}3{\Cal H}+\ln I<\ln 2,
\end{equation}
then this motion is a simple relative choreography~$($see
Fig.~{\rm\ref{plane.eps})}.\goodbreak
\end{teo}

\proof
 Since in the case of three vortices, the reduced
system~\eqref{D3} has one degree of freedom, all~$M_{ij}(t)$ are periodic
functions of the same period~$T$. It can be easily shown that under the
restriction~\eqref{D5}, the orientation of the vortex triangle is unchanged,
and there are times~$t_1$ and~$t_2$ such that~$M_{23}(t_1)=M_{13}(t_2)$.
(If~\eqref{D5} is not met, then~$\forall\,t$ $\exists\,k$ such
that~$M_{ij}(k)<M_{ik}(t)$ and~$M_{ij}(k)<M_{jk}(t)$.) Moreover, since
with fixed~$H$, $I$ all~$M_{ij}$ are expressed through one of them (for
example,~$M_{12}$), the following relations also hold true:
$$
M_{13}(t_1)=M_{12}(t_2),\quad M_{12}(t_1)=M_{23}(t_2).
$$
Since the equations are invariant under cyclic permutation of the vortices
(as implied by equality of the intensities), we find that
$$
t_1-t_2=\frac T3n,\quad n\in\mathbb{Z}.
$$
Since the evolution equation for~$M_{ij}$ is of the first order, we
conclude that
$$
M_{12}(t)=M_{23}\left(t+\frac
T3\right)=M_{13}\left(t+\frac{2T}3\right)=f(t),
$$
or
$$
M_{12}(t)=M_{23}\left(t+\frac{2T}3\right)=M_{13}\left(t+\frac T3\right)=f(t),
$$
where $f(t)$ is some~$T$-periodic function.

Substituting this into~\eqref{D2}, we obtain
$$
z_k(t)=u\left(t+\frac{k-1}3T\right)e^{i\Omega t},\quad k=1,2,3,
$$
where $u(t)$ is a~$T$-periodic complex-valued function, determining the
same curve, along which the vortices move, in the frame of reference
rotating with angular velocity~$\Omega$. The phase portrait of the reduced
system of the~3-vortex problem and the corresponding relative
choreographies are given in Fig.~\ref{plane.eps}. \qed

\fig<bb=0 0 120.2mm 63.6mm>{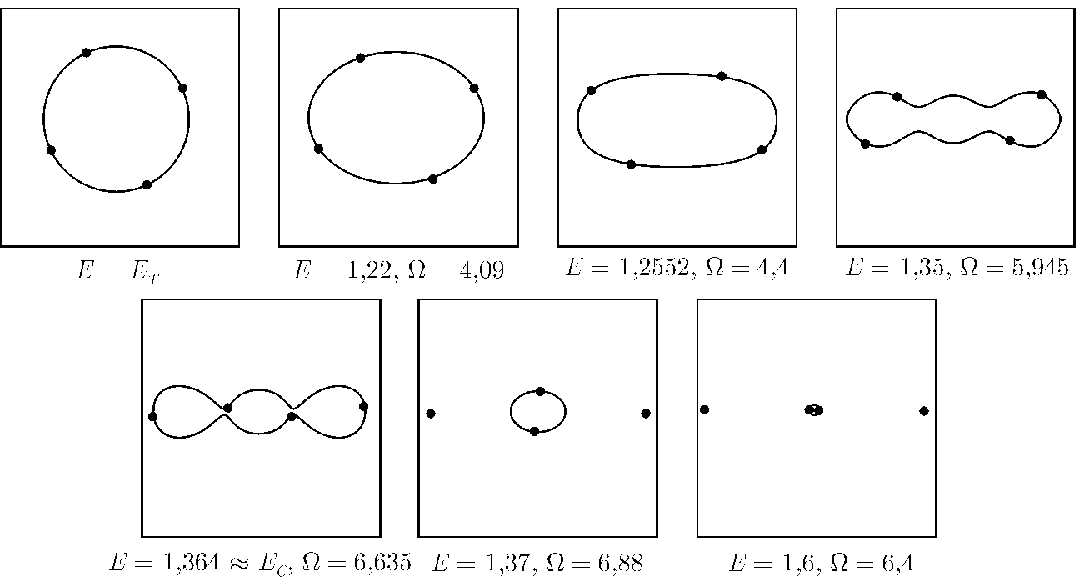}[Relative choreographies in
the~4-vortex problem.]

The~4-vortex problem has an exceptional solution, given by quadratures~---
Goryachev's solution, where the vortices form a parallelogram at each
instant of time~\cite{GoryachevSbornik}. As in the case of the~3-vortex
problem, it is quite easy to show that

\begin{teo}[\cite{bmk-1}]\label{teo2}
If in the 4-vortex problem, the vortices $($of equal intensity\/$)$ form a
centrally symmetric configuration $($a parallelogram\/$)$, while the
constants~$\Cal H$ and $I$ satisfy the inequality
$$
-\ln2<\frac{2\pi}3{\Cal H}+\ln I<-\ln\frac{144}5,
$$
then the motion is a simple relative choreography.
\end{teo}

The corresponding relative choreographies are shown in
Fig.~\ref{par_gram.eps}.

\begin{rem*}
The physical meaning of the inequality is as follows:  when~$I$ is fixed,
the type of the motion in the 3-vortex problem and in the case of Goryachev's
solution is changed at energy values corresponding to the Thomson and collinear
configurations.
\end{rem*}

\paragraph{New periodic solution in the~4-vortex problem.} Now we show
that, aside from the choreography just described, the~4-vortex problem has
at least one more choreography (different from Goryachev's solution).
Consider the vicinity of Thomson's solution, i.\,e. the motion, where the
vortices are located at the vertices of a square and rotate uniformly
about the vorticity center~\cite{Tomson}. It is obvious that in the case
of the reduced system with two degrees of freedom~\eqref{D4}, Thomson's
solution is represented by a fixed point (more precisely, by six points
corresponding to various arrangements of vortices at
square's vertices). We consider one of the arrangenents (the rest are completely
identical) with  coordinates
$$
G=0,\quad H=\frac12,\quad h=\frac\pi2.
$$
\fig<bb=0 0 83.3mm 64.4mm>{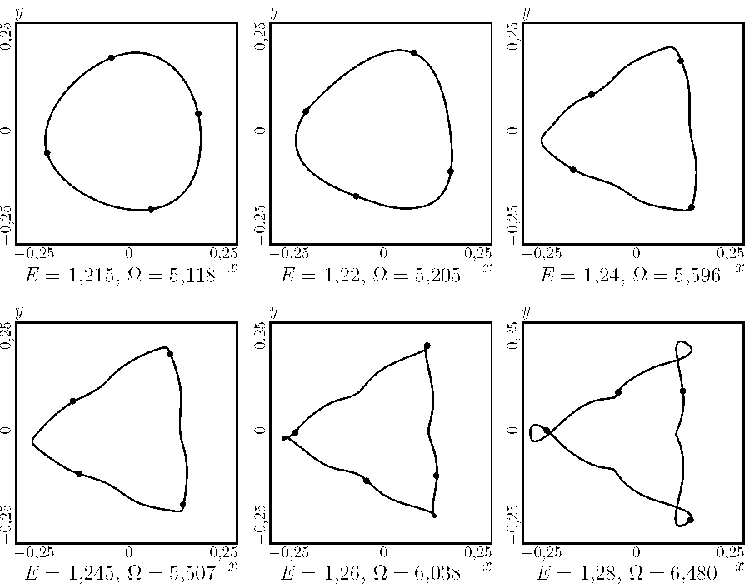}[\label{trapez.eps} Relative
choreographies corresponding to the new periodic solution in the
three-body problem.] Let us find the normal form of the reduced system's
\eqref{D4} Hamiltonian near this point. To do that, we start with a
canonical change of variables:
$$
G=\frac{x^2+X^2}2,\quad g=\arctg\frac xX,\quad H=\frac12+8^{-1/4}Y,\quad
h=\frac\pi2+8^{1/4}y;
$$
Expanding the Hamiltonian in a  series up to and including quadratic terms, we find
\begin{gather}
\label{E1}
{\Cal H}=\frac{\ln2}\pi+{\Cal H}_2+{\Cal H}_r,\\
{\Cal H}_2=\frac1{4\pi}\left(3(x^2+X^2)+2\sqrt2(y^2+Y^2)\right),
\notag
\end{gather}
where the expansion of~${\Cal H}_r$ starts with the third-order terms.
Thus, the Hamiltonian~${\Cal H}_2$ defines an integrable system with two
incommensurable frequencies and has precisely two non-degenerate periodic
solutions on each energy level~${\Cal H}_2=h_2=\const$. These solutions
are given by
\begin{gather}
\label{E2}
x=X=0,\quad y^2+Y^2=\frac\pi{\sqrt2}h_2,\\
\label{E3}
y=Y=0,\quad x^2+X^2=\frac{4\pi}3h_2.
\end{gather}

\fig<bb=0 0 83.5mm 64.4mm>{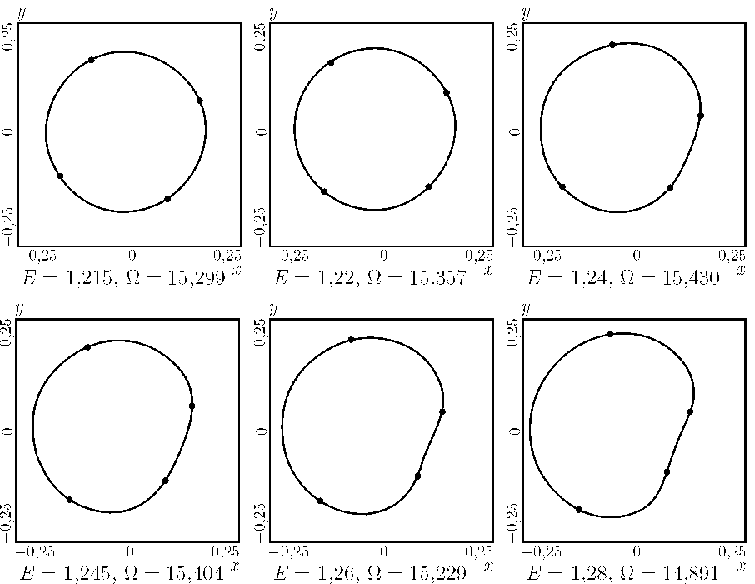}[\label{four.eps}Relative
choreographies corresponding to the new periodic solution in the
three-body problem in the frame of reference different from that used in
Fig.~\ref{trapez.eps}.]

According to the Lyapunov theorem~\cite{Zigel}, these solutions are preserved  under perturbations, hence, the complete system~\eqref{E1} in the
vicinity of the fixed point also has two non-degenerate periodic solutions
on each energy level. It is easy to show that one the solutions (corresponding
to~\eqref{E2}) is identical to Goryachev's solution ---
during their motion the vortices are located at the vertices of a parallelogram.
At the same time, the other
solution~\eqref{E3} does not have such a simple geometric interpretation.

Since  equations \eqref{bmk-eq-1} are invariant under a cyclic
permutation of the vortices~$\sigma_c(z_1,\,z_2,\,z_3)=(z_3,\,z_1,\,z_2)$
and the eigenvalues of the Hamiltonian~${\Cal H}_2$ are different, it
is easy to show that both  periodic solutions are also
invariant under~$\sigma_c$. Thus, according to Proposition~\ref{pro3}, in
the appropriate frame of reference, all the vortices move along the same
curve, i.\,e. both solutions correspond to simple relative choreographies.
Figure~\ref{trapez.eps} shows the relative choreographies that correspond
to the new periodic solution of the reduced system~\eqref{D4}.

\paragraph{Relative and absolute choreographies.}
Generally, for each periodic solution (of period~$T$) of the reduced
system~\eqref{D3}, \eqref{D4}, it is possible to specify a countable set
of rotating frames of reference, where the vortices move along closed
curves. Indeed, the trajectories will remain closed in a frame rotating with velocity
\begin{equation}
\label{F1}
\Omega_a'=\Omega_a+\frac pq\frac{2\pi}T,\quad p,q\in\mathbb{Z}
\end{equation}
Nevertheless, the change~\eqref{F1} with
arbitrary~$p$ and~$q$ does not preserve the connectedness of the
trajectories, i.\,e. in the general case, after moving to the frame of
reference, rotating with velocity~$\Omega_a'$, a simple relative
choreography decomposes into separate closed curves, along which the
vortices move. To preserve the connectedness, the following criterion
should be met.

\begin{pro}
Let a periodic solution $($of period~$T)$ of the reduced system describe a
connected relative choreography, the rotational velocity of the frame of
reference being~$\Omega_a$, while the period of the motion of the vortices
along the corresponding common curve being equal to~$mT$. Then, if
\begin{equation}
\label{F2}
mp=knq,
\end{equation}
where $n$ is the number of vortices, and~$k\in\mathbb{Z}$ is an arbitrary
integer, the transformation~\eqref{F1} results in a connected
choreography.
\end{pro}

\proof
 For the solution in question, the absolute coordinates
of the vortices can be presented in the form:
\begin{equation}
\label{F3}
z_k(t)=u\left(t+\frac{k-1}nmT\right)e^{i\Omega_at},
\end{equation}
where $u(t)=u(t+mT)$ is a periodic complex-valued function (of
period~$mT$).

Solving for~$\Omega_a$ from~\eqref{F1} and substituting into~\eqref{F3}, we get
\begin{equation}
\label{F4}
z_k(t)=u\left(t+\frac{k-1}nmT\right)e^{-i\frac{2\pi}T\frac
pqt}e^{i\Omega_a't}=u_k(t)e^{i\Omega_a't}.
\end{equation}
If all the vortices move along the same curve, then their coordinates in
the rotating (with velocity~$\Omega_a'$) frame of reference are equal
to~$u_k(t)$ and satisfy~$u_{k+1}(t)=u_k\left(t+\frac{mT}n\right)$, whence,
taking~\eqref{F4} into account, we obtain~\eqref{F2}. \qed

The relation~\eqref{F2} is a sufficient but not necessary condition for the
choreography  to be connected. If the curve along which the vortices move
has additional symmetries, then, besides the~$p,\,q$ that meet the
condition~\eqref{F2}, there are more velocities of the form~\eqref{F1},
which result in connected choreographies (see below, for Goryachev's
solution).

\fig<bb=0 0 92.1mm 80.3mm>{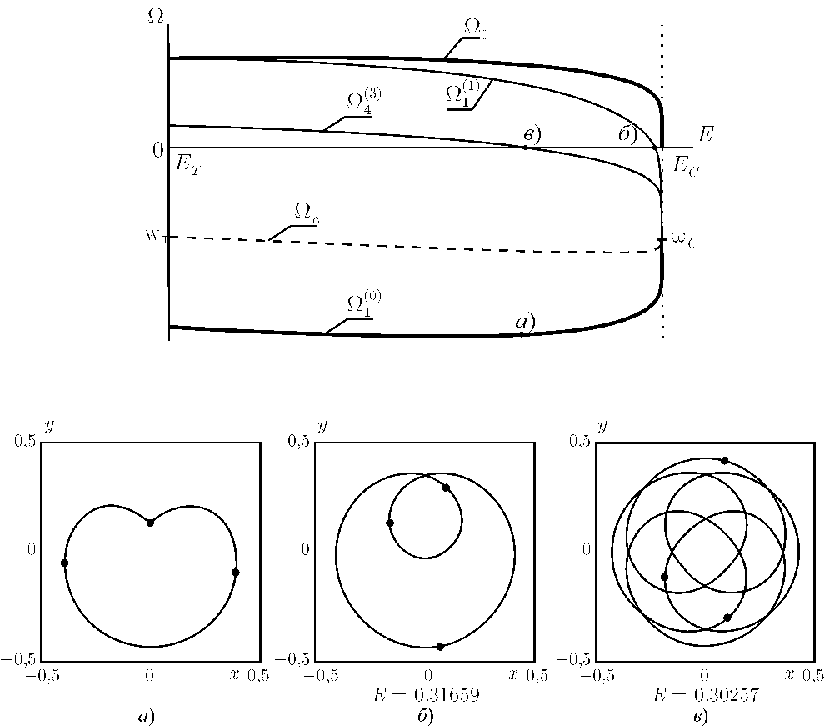}[\label{five.eps} The rotational
angular velocities,~$\Omega_1^{(1)}$ and~$\Omega_4^{(3)}$, of relative
choreographies are presented as functions of energy. The points where the
graphs intersect the~$Ox$ axis correspond to the absolute choreographies
of the three vortices shown in the bottom (Figs.~b and~c). The heavy lines
denote the basic angular velocity~$\Omega_1^{(0)}$, corresponding to the
simplest relative choreography (shown in Fig.~a), and the frequency of the
periodic solution of the reduced
system~\eqref{D3},~$\Omega_0=\frac{2\pi}{T}$.
]

It is interesting to note that using the transformation~\eqref{F1}, some choreographies
can be ``disentangled''~--- for example,
Fig.~\ref{four.eps} shows choreographies, corresponding to the periodic
solution~\eqref{E3}, in the frame of reference, different from that used
in Fig.~\ref{trapez.eps}, their rotational velocities differ by
$$
\Omega'-\Omega=\frac43\frac{2\pi}T.
$$

As it was shown above (see Proposition~\ref{pro3}), if, for a relative
choreography, the period~$T$ of the reduced system's solution is
commensurable with the rotational period~$T_a=\frac{2\pi}{\Omega_a}$ of
the frame of reference, then in a fixed frame of reference, all the
vortices move along closed (and, usually, different) curves.

Let us consider in greater detail the existence of absolute choreographies in the three- and
four-vortex problems. According to what was said above, any
relative choreography, corresponding to the periodic solution (of
period~$T$) of the reduced system (see~\eqref{D3},~\eqref{D4}), closes in
time~$mT,\,m\in\mathbb{N}$. During this time interval, the vortices pass
one and the same relative configuration~$m$ times. Let~$\Omega_m^{(k)}$ be the
rotational  velocities of the frames of reference related to these
choreographies.

\fig<bb=0 0 83.8mm 71.6mm>{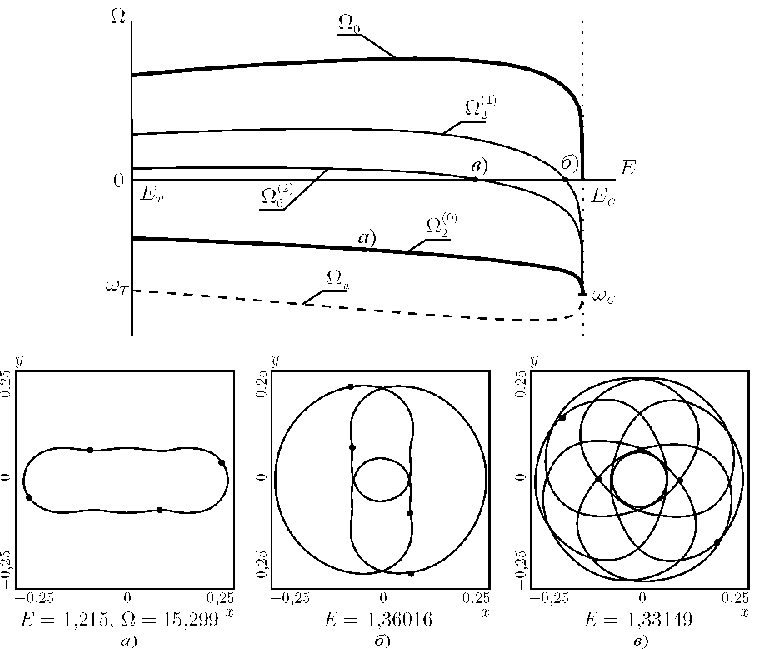}[The rotational angular
velocities,~$\Omega_2^{(1)}$ and~$\Omega_6^{(2)}$, of relative
choreographies are presented as functions of energy. The points where the
graphs intersect the~$Ox$ axis correspond to the absolute choreographies
of the four vortices shown in the bottom (Figs.~b and~c). The heavy lines
denote the basic angular velocity~$\Omega_2^{(0)}$, corresponding to the
simplest relative choreography (shown in Fig.~a), and the frequency of the
reduced system~\eqref{D4},~$\Omega_0=\frac{2\pi}{T}$.
]

As it was shown above, all the solutions of the reduced three-vortex
system, for a fixed~$D$ and~$E_T<E<E_C$ (where~$E_T$ and~$E_C$ are the
energies corresponding to the Thomson and collinear configurations),
describe connected relative choreographies. Moreover, there is a frame
of reference, where the choreography closes in (the shortest possible)
time~$T$ (see Fig.~\ref{five.eps}); the corresponding angular velocity is
denoted as~$\Omega_1^{(0)}$, its graph is shown in Fig.~\ref{five.eps}.
The angular velocities of the other relative connected choreographies are now
given by
\begin{equation}
\label{F5}
\Omega_m^{(k)}(E)=\Omega_1^{(0)}(E)+\frac{3k}{m}\Omega_0(E),\text{ where }
m\in\mathbb{N},\,k\in\mathbb{Z};
\end{equation}
here~$3k$ and $m$ are coprime numbers, and~$\Omega_0(E)=\frac{2\pi}{T(E)}$.
The velocity~$\Omega_m^{(k)}$ corresponds to the choreography that closes in
time~$mT$.

Obviously, the absolute choreographies are defined by the solutions of the
equation
\begin{equation}
\label{F6}
\Omega_m^{(k)}(E)=0,
\end{equation}
where $k,\,m$ are fixed, while~$E$ is unknown. Figure~\ref{five.eps} shows
the graphs of  velocities~$\Omega_m^{(k)}(E)$ together with the
solutions of  equation~\eqref{F6}, as well as the corresponding
absolute choreographies (the simplest choreographies in a three-vortex
system). Generally, there is a countable set of absolute choreographies
with different~$m,\,k$.

Indeed, consider the function
$$
f_a(E)=\Omega_1^{(0)}(E)+a\Omega_0(E),\q a\in\mathbb{R},
$$
for which the equality~$f_a(E_C)=\omega_C<0$ holds (see
Fig.~\ref{five.eps}). Since~$\Omega_0(E)>0$, there exists  a
number~$a_*$ such that when~$a>a_*$, the function~$f_a(E)$ has at least
one zero in the interval~$[E_T,\,E_C]$. It is clear that the
interval~$[a_*,\,+\infty)$ contains an infinite number of rationals of the
form~$a=\frac{3k}{m}$, where~$3k$ and~$m$ are coprimes. The dotted line in
Fig.~\ref{five.eps} shows the curve~$f_{a^*}(E)$ and the relative
choreography's rotational angular velocity~$\Omega_a$, calculated
from~\eqref{Om1}. This choreography is the simplest disconnected
choreography and is remarkable for the fact that the values at the
ends of the interval,~$\Omega_a(E_T)$ and $\Omega_a(E_C)$, are
equal to the rotational angular velocities of the Thomson and collinear
configurations.

For Goryachev's solution in the four-vortex system, the reasoning is
similar but slightly modified. First of all, one can show that the
simplest connected choreography closes in time~$2T$, while the vortices in
this case pass one and the same relative configuration twice (i.\,e.
velocities~$\Omega_1^{(k)}$ correspond to disconnected choreographies).
The graph of one of the corresponding angular velocities, which we denote
as~$\Omega_2^{(0)}$, is given in Fig.~\ref{fouromega.eps}. In this case,
due to the symmetry of the curve related to the
choreography~$\Omega_2^{(0)}$, the frequencies corresponding to connected
choreographies must satisfy  the relation different from~\eqref{F5},
\begin{equation}
\label{F7}
\Omega_{2m}^{(k)}(E)=\Omega_2^{(0)}(E)+\frac{k}{m}\Omega_0(E),\text{ where
} m\text{ is odd},\,k\in\mathbb{Z};
\end{equation}
here~$k$ and $m$ are coprimes, and~$\Omega_0(E)=\frac{2\pi}{T(E)}$,
where~$T$ is the period of Goryachev's solution to the reduced
system~\eqref{D4}. This choreography closes in time~$2mT$.

As above, the equation
$$
\Omega_{2m}^{(k)}(E)=0
$$
defines the absolute choreographies. Similar to the three-vortex problem,
one can show that there is a countable set of absolute choreographies,
described by Goryachev's solution with different~$m$ and~$k$.

\paragraph{Stability.} \mbox{As a conclusion, let us discuss the stability
of the specified periodic solutions.}

Since the three-vortex problem is integrable, all the solutions of the
reduced system~\eqref{D3} are periodic and stable. Yet it is easy to show
that any (absolute or relative) choreography in this problem is neutrally
stable under perturbations of the vortices' positions in the absolute
space.

In the four-vortex problem, due to its non-integrability, the relative
choreographies can be (exponentially) unstable. Yet, if a periodic
solution of the reduced system~\eqref{D4} is stable, the corresponding
choreographies are also neutrally stable in the absolute space. The
numerical analysis of the multipliers of the periodic solutions that
describe choreographies presented in Fig.~\ref{fouromega.eps}~b,~c shows
that they are (exponentially) unstable.

\medskip

We gratefully acknowledge support of the Russian Foundation for Basic Research
(grant \No~04\205\264367), the Foundation for Leading Scientific Schools
(grant \No~136.2003.1) and CRDF (grant \No~RU-M1-2583-MO-04).

\end{document}